\documentclass[11pt,draftclsnofoot,journal,onecolumn]{IEEEtran}
\usepackage{algorithm}
\usepackage{algorithmic}
\usepackage[dvips]{graphicx}
\usepackage{epsfig}
\usepackage{graphics}
\usepackage{float}
\usepackage{times}
\usepackage{color}
\usepackage{mathrsfs}
\usepackage{amsmath}
\usepackage{amssymb}
\usepackage{cite}
\usepackage{subfigure}
\usepackage{subfigure}

\begin{document}

\title{Extensions to the Theory of Widely Linear Complex Kalman Filtering}
\author{Wenbing Dang, \textit{Student Member}, \textit{IEEE}, and Louis L. Scharf, \textit{Life Fellow}, \textit{IEEE}}\maketitle
\let\thefootnote\relax\footnotetext{This work is supported by the National Science Foundation under grants CCF-1018472 and CCF-0916314, and the Air Force Office of Scientific Research under contract FA9550-10-1-0241.}
\footnotetext{W. Dang is with the Department of Electrical and Computer Engineering, Colorado State University, Fort Collins, CO 80523, USA, Phone: (970)
817-0249 (email: dwb87514@gmail.com).}
\footnotetext{L. L. Scharf is with the Departments of Mathematics and Statistics, Colorado State
University, Fort Collins, CO 80523, USA, Phone: (970) 491-2979, Fax: (970) 491-2249 (e-mail: Louis.Scharf@colostate.edu).}
\begin{abstract}
For an improper complex signal $\mathbf{x}$, its complementary covariance $E\mathbf{x}\mathbf{x}^T$ is not zero and thus it carries useful statistical information about $\mathbf{x}$. Widely linear processing exploits Hermitian \textit{and} complementary covariance to improve performance. In this paper we extend the  existing theory of widely linear complex Kalman filters (WLCKF) and unscented WLCKFs \cite{Mandic_book}. We propose a WLCKF which can deal with more general dynamical models of complex-valued states and measurements than the WLCKFs in \cite{Mandic_book}. The proposed WLCKF has an equivalency with the corresponding dual channel real KF. Our analytical and numerical results show the performance improvement of a WLCKF over a complex Kalman filter (CKF) that does not exploit complementary covariance. We also develop an unscented WLCKF which uses modified complex sigma points. The modified complex sigma points preserve complete first and second moments of complex signals, while the sigma points in \cite{Mandic_book} only carry the mean and Hermitian covariance, but not complementary covariance of complex signals.
\end{abstract}
\section{Introduction}
Complex signals are ubiquitous in science and engineering, arising as they do as complex representations of two real channels or of two-dimensional fields. Consider a zero mean complex random vector $\mathbf{x}$. The usual covariance matrix defined as $E\mathbf{x}\mathbf{x}^H$ describes its Hermitian second order covariance. But when $\mathbf{x}$ and its complex conjugate $\mathbf{x}^*$ are correlated, the complementary covariance matrix $E\mathbf{x}\mathbf{x}^T$ does not vanish, so it carries useful second order information about the complex random vector $\mathbf{x}$. We call a complex random vector proper as long as its complementary covariance matrix vanishes and improper otherwise. Proper complex vectors have a statistical description similar to real vectors, but improper random vectors do not. A comprehensive second order analysis of improper random vectors and processes is considered in \cite{Picinbono1, Peter_secondorder, book_PJ&LL, Koivunen, Complex_MMSE}.

For any improper random vector $\mathbf{x}$, for which $\mathbf{x}$ is correlated with its complex conjugate $\mathbf{x}^*$, intuition suggests that a good estimator of $\mathbf{x}$ should depend on $\mathbf{x}^*$. This requires a methodology of \emph{widely linear processing} instead of strictly linear processing \cite{Picinbono1}. For random complex signals, the merit of widely linear processing has been exploited in various papers on estimation \cite{book_PJ&LL, Complex_MMSE}, filtering \cite{Peter_secondorder, book_PJ&LL, Mandic_book}, detection \cite{ML_det, Improper_det}, and equalization \cite{WL_equal}. It turns out that widely linear processing brings improvement in performance over strictly linear processing \cite{book_PJ&LL, WLproc_Mandic} when there is complementary covariance to be exploited.

In the past few decades the reasoning of the Kalman filter \cite{KF} has been modified to apply to nonlinear problems, producing Extended Kalman filters \cite{DAUM_NKF} and Unscented Kalman filters \cite{UKF_Julier1}. The motivation of this paper is to make use of widely linear processing to develop novel complex Kalman filters and their nonlinear versions for improper complex states. We show that for improper complex states, complementary covariance matrices may be used to create widely linear complex KFs (denoted WLCKFs) and Unscented WLCKFs. The key contributions of this paper are as follows:
\begin{itemize}
  \item From a \emph{linear} real dual channel dynamical model we derive an equivalent \emph{widely linear} complex single channel dynamical model, where the updates of random states and measurements depend on both states and noises and their conjugates. For the complex model we derive a WLCKF which is equivalent to the conventional KF for the dual channel model. The WLCKFs proposed in \cite{Mandic_book} consider special dual channel problems and their corresponding complex dynamical models. In these complex models the updates of complex random states and measurements do not depend on the conjugates of states and noises.
  \item We compare the performance between the WLCKFs and conventional KFs. Our analytical and numerical results show that for some special distributions of states and noises, the MSE of the WLCKF is significantly smaller than the MSE of a CKF that does not exploit non-zero complementary covariance.
  \item For dynamical models with complex nonlinear state and measurement equations, we develop an Unscented WLCKF for which a systematic paradigm to construct \emph{modified} complex sigma points is studied. The property of modified sigma points is that they preserve the complete first and second order statistical information of complex random vectors. The WLCKF of \cite{Mandic_book} uses sigma points that only preserve the mean and Hermitian covariance, but not the complementary covariance of states.
\end{itemize}

\section{Brief Review of Complex Random Vectors}
Let $\Omega$ be the sample space of a random experiment that generates two channels of real signals $\mathbf{u}, \mathbf{v}\in \mathbb{R}^n$ defined on $\Omega$. From this we construct the \emph{real composite} random vector $\mathbf{z}\in \mathbb{R}^{2n}$ as $\mathbf{z}^T=[\mathbf{u}^T, \mathbf{v}^T]$, and the \emph{complex} random vector $\mathbf{x}\in\mathbb{C}^n$, obtained by composing $\mathbf{u}$ and $\mathbf{v}$ into its real and imaginary parts:
\begin{equation}
\mathbf{x} = \mathbf{u}+j\mathbf{v} .
\end{equation}
The \emph{complex} \emph{augmented} random vector $\underline{\mathbf{x}}$ corresponding to $\mathbf{x}$ is defined as
\begin{equation}
\underline{\mathbf{x}}^T = [\mathbf{x}^T\ \mathbf{x}^H] .
\end{equation}
From here the complex augmented random vector will always be underlined. It's easy to check that the real composite vector $\mathbf{z}$ and the complex augmented vector $\underline{\mathbf{x}}$ are related as
\begin{equation}\label{tranf}
\underline{\mathbf{x}} = \mathbf{T}_{n}\mathbf{z}.
\end{equation}
The real-to-complex transformation $\mathbf{T}_n$ is
\begin{equation}
\mathbf{T}_n = \begin{bmatrix}\mathbf{I} & j\mathbf{I} \\ \mathbf{I} & -j\mathbf{I}\end{bmatrix} ,
\end{equation}
which is unitary within a factor of 2:
\begin{equation}
\mathbf{T}_n\mathbf{T}_n^H = \mathbf{T}_n^H\mathbf{T}_n = 2\mathbf{I} .
\end{equation}
In fact, it is equation (\ref{tranf}) that governs the equivalence between dual channel filtering for $\mathbf{z}$ and complex filtering for $\mathbf{x}$.

The augmented mean vector of the complex random vector $\mathbf{x}$ is
\begin{equation}
\underline{\boldsymbol{\mu}}_{\mathbf{x}} = E\underline{\mathbf{x}}= [\boldsymbol{\mu}_x^T \ \boldsymbol{\mu}_x^H]^T = [ \boldsymbol{\mu}_u^T + j\boldsymbol{\mu}_v^T \ \boldsymbol{\mu}_u^T - j\boldsymbol{\mu}_v^T]^T = \mathbf{T}\boldsymbol{\mu}_{z},
\end{equation}
and the augmented covariance matrix of $\mathbf{x}$ is
\begin{equation}\label{agcov}
\underline{\mathbf{R}}_{xx} = E(\underline{\mathbf{x}}-\underline{\boldsymbol{\mu}}_x)
(\underline{\mathbf{x}}-\underline{\boldsymbol{\mu}}_x)^H =
\begin{bmatrix}\mathbf{R}_{xx} & \widetilde{\mathbf{R}}_{xx} \\ \widetilde{\mathbf{R}}_{xx}^* & \mathbf{R}_{xx}^*\end{bmatrix} = \mathbf{T}\mathbf{R}_{zz}\mathbf{T}^H,
\end{equation}
where the matrix $\mathbf{R}_{xx} = E(\mathbf{x}-\boldsymbol{\mu}_x)
(\mathbf{x}-\boldsymbol{\mu}_x)^H$ is the conventional Hermitian covariance matrix,
and the matrix $\widetilde{\mathbf{R}}_{xx} = E(\mathbf{x}-\boldsymbol{\mu}_x)
(\mathbf{x}-\boldsymbol{\mu}_x)^T$ is the complementary covariance matrix

\emph{Definition 1:}
If the complementary covariance matrix $\widetilde{\mathbf{R}}_{xx}$ is zero, then $\mathbf{x}$ is called proper; otherwise $\mathbf{x}$ is improper.

The random vector $\mathbf{x}=\mathbf{u}+j\mathbf{v}$ is proper if and only if $\mathbf{R}_{uu}=\mathbf{R}_{vv}$ and $\mathbf{R}_{uv}=-\mathbf{R}_{uv}^T$, where $\mathbf{u}$ and $\mathbf{v}$ are the real and imaginary parts of $\mathbf{x}$ respectively.

\section{Dual Channel Real and Widely-Linear Complex Kalman Filter}
Start with two \textit{real} channels worth of random states $\mathbf{u}_t, \mathbf{v}_t\in \mathbb{R}^{n}$. Denote $\mathbf{z}_t^T=[\mathbf{u}_t^T\ \mathbf{v}_t^T]$ as the corresponding real \textit{composite} state. Suppose the composite state and measurement equations are
\begin{equation}\label{ct_kf1}
\mathbf{z}_{t} = \begin{bmatrix}\mathbf{u}_{t} \\ \mathbf{v}_{t}\end{bmatrix}=\mathbf{E}\mathbf{z}_{t-1}+\mathbf{F}\boldsymbol{\omega}_{t-1}
=\begin{bmatrix}\mathbf{E}_{11} & \mathbf{E}_{12} \\ \mathbf{E}_{21} & \mathbf{E}_{22}\end{bmatrix}\begin{bmatrix} \mathbf{u}_{t-1}\\ \mathbf{v}_{t-1}\end{bmatrix}+\begin{bmatrix}\mathbf{F}_{11} & \mathbf{F}_{12} \\ \mathbf{F}_{21} & \mathbf{F}_{22}\end{bmatrix}\begin{bmatrix}\boldsymbol{\mu}_{t-1} \\ \boldsymbol{\sigma}_{t-1}\end{bmatrix},\ t=1,2,...,
\end{equation}
and
\begin{eqnarray}\label{ct_kf2}
&&\boldsymbol{\psi}_t = \begin{bmatrix}\boldsymbol{\xi}_{t} \\ \boldsymbol{\kappa}_{t}\end{bmatrix}=\mathbf{G}\mathbf{z}_t+\boldsymbol{\eta}_t
=\begin{bmatrix}\mathbf{G}_{11} & \mathbf{G}_{12} \\ \mathbf{G}_{21} & \mathbf{G}_{22}\end{bmatrix}\begin{bmatrix}\mathbf{u}_{t} \\ \mathbf{v}_{t}\end{bmatrix}+ \begin{bmatrix}\boldsymbol{\rho}_{t} \\ \boldsymbol{\phi}_{t}\end{bmatrix},\ t=0,1,...,
\end{eqnarray}
where $\boldsymbol{\omega}_t^T=[\boldsymbol{\mu}_{t}^T\ \boldsymbol{\sigma}_t^T]$ and $\boldsymbol{\eta}_t^T=[\boldsymbol{\rho}_{t}^T\ \boldsymbol{\phi}_t^T]$ are the composite real driving and measurement noises, and $\boldsymbol{\psi}_t^T=[\boldsymbol{\xi}_t^T\ \boldsymbol{\kappa}_t^T]$ is the composite measurement. This dynamical model allows the states and measurements on the respective real channels to be arbitrarily coupled. For the real composite vectors $\mathbf{z}_t$, $\boldsymbol{\omega}_t$, $\boldsymbol{\eta}_t$, and $\boldsymbol{\psi}_t$, establish their complex augmented representations as $\underline{\mathbf{x}}_t = [\mathbf{x}_t^T\  \mathbf{x}_t^H]^T = \mathbf{T}\mathbf{z}_t$, $\underline{\mathbf{w}}_t = [\mathbf{w}_t^T\  \mathbf{w}_t^H]^T=\mathbf{T}\boldsymbol{\omega}_t$, $\underline{\mathbf{y}}_t = [\mathbf{y}_t^T\  \mathbf{y}_t^H]^T = \mathbf{T}\boldsymbol{\psi}_t$, and $
\underline{\mathbf{n}}_t = [\mathbf{n}_t^T\  \mathbf{n}_t^H]^T = \mathbf{T}\boldsymbol{\eta}_t$. Then the resulting augmented complex state and measurement equations are
\begin{equation}\label{aug_kf1}
\underline{\mathbf{x}}_{t} = \underline{\mathbf{A}}\underline{\mathbf{x}}_{t-1} + \underline{\mathbf{B}}\underline{\mathbf{w}}_{t-1},\ t = 1,2,...,
\end{equation}
\begin{equation}\label{aug_kf2}
\underline{\mathbf{y}}_{t} = \underline{\mathbf{C}}\underline{\mathbf{x}}_{t} + \underline{\mathbf{n}}_{t},\ t = 0, 1,...,
\end{equation}
where the augmented matrices $\underline{\mathbf{A}}$, $\underline{\mathbf{B}}$, and $\underline{\mathbf{C}}$ are
\begin{equation}\label{matrx}
\underline{\mathbf{A}}=\frac{1}{2}\mathbf{T}\mathbf{E}\mathbf{T}^{H}=\begin{bmatrix}\mathbf{A}_1 & \mathbf{A}_2 \\ \mathbf{A}_2^* & \mathbf{A}_1^*\end{bmatrix}, \underline{\mathbf{B}}=\frac{1}{2}\mathbf{T}\mathbf{F}\mathbf{T}^{H}=\begin{bmatrix}\mathbf{B}_1 & \mathbf{B}_2 \\ \mathbf{B}_2^* & \mathbf{B}_1^*\end{bmatrix}, \underline{\mathbf{C}}=\frac{1}{2}\mathbf{T}\mathbf{G}\mathbf{T}^{H}=\begin{bmatrix}\mathbf{C}_1 & \mathbf{C}_2 \\ \mathbf{C}_2^* & \mathbf{C}_1^*\end{bmatrix}.
\end{equation}
Suppose the initial state has mean $E\underline{\mathbf{x}}_0=\mathbf{0}$, and augmented covariance
\begin{equation}\label{init}
E\underline{\mathbf{x}}_0\underline{\mathbf{x}}_0^H = \begin{bmatrix}E\mathbf{x}_0\mathbf{x}_0^H & E\mathbf{x}_0\mathbf{x}_0^T \\ E\mathbf{x}_0^*\mathbf{x}_0^H & E\mathbf{x}^*_0\mathbf{x}_0^T\end{bmatrix} = \begin{bmatrix}\boldsymbol{\Pi}_0 & \widetilde{\boldsymbol{\Pi}}_0 \\ \widetilde{\boldsymbol{\Pi}}_0^* & \boldsymbol{\Pi}_0^*\end{bmatrix} = \underline{\boldsymbol{\Pi}}_0.
\end{equation}
Using the representation advocated in \cite{book_est}, the augmented second-order characterization of $(\underline{\mathbf{x}}_0,\underline{\mathbf{u}}_t,\underline{\mathbf{n}}_t)$ is given by
\begin{equation}
E\begin{bmatrix}\underline{\mathbf{x}}_0 \\ \underline{\mathbf{w}}_n \\ \underline{\mathbf{n}}_n \end{bmatrix}\begin{bmatrix}\underline{\mathbf{x}}_0^H & \underline{\mathbf{w}}_m^H & \underline{\mathbf{n}}_m^H & \underline{\mathbf{1}}^H \end{bmatrix} = \begin{bmatrix}\underline{\boldsymbol{\Pi}}_0 & \mathbf{0} & \mathbf{0} &\mathbf{0} \\ \mathbf{0} & \delta_{nm}\underline{\mathbf{Q}} & \delta_{nm}\underline{\mathbf{S}} &\mathbf{0} \\ \mathbf{0} & \delta_{nm}\underline{\mathbf{S}}^H & \delta_{nm}\underline{\mathbf{R}}& \mathbf{0}\end{bmatrix} , m,n\geq 0.
\end{equation}
We further assume that for $n\geq m$, $E\underline{\mathbf{w}}_n\underline{\mathbf{x}}_m^H = \mathbf{0}$ and $E\underline{\mathbf{n}}_n\underline{\mathbf{x}}_m^H = \mathbf{0}$, and for $n > m$, $E\underline{\mathbf{w}}_n\underline{\mathbf{y}}_m^H = \mathbf{0}$ and $E\underline{\mathbf{n}}_n\underline{\mathbf{y}}_m^H = \mathbf{0}$.
This is the same setup as that of the usual Kalman filter, but with covariances augmented to account for non-zero complementary covariance.

Suppose the LMMSE estimator of $\underline{\mathbf{x}}_{t-1}$ from measurements $\underline{\mathbf{Y}}_{t-1}^T =(\underline{\mathbf{y}}_{1}^T, \ldots,\underline{\mathbf{y}}_{t-1})^T$ is $\hat{\underline{\mathbf{x}}}_{t-1|t-1}$. Then the LMMSE prediction of $\underline{\mathbf{x}}_{t}$ from $\underline{\mathbf{Y}}_{t-1}^T$ is
\begin{equation}\label{wlkf_bg}
\hat{\underline{\mathbf{x}}}_{t|t-1}=\underline{\mathbf{A}}\hat{\underline{\mathbf{x}}}_{t-1|t-1},
\end{equation}
and the prediction of $\underline{\mathbf{y}}_{t}$ from $\underline{\mathbf{Y}}_{t-1}^T$ is
\begin{equation}
\hat{\underline{\mathbf{y}}}_{t|t-1}=\underline{\mathbf{C}}\hat{\underline{\mathbf{x}}}_{t|t-1}.
\end{equation}

Given the error covariance matrix $\underline{\mathbf{P}}_{t-1|t-1}$ for $\hat{\underline{\mathbf{e}}}_{t-1|t-1}=\hat{\underline{\mathbf{x}}}_{t-1|t-1}-\underline{\mathbf{x}}_{t-1}$, the error covariance matrix $\underline{\mathbf{P}}_{t|t-1}$ for $\hat{\underline{\mathbf{e}}}_{t|t-1}=\hat{\underline{\mathbf{x}}}_{t|t-1}-\underline{\mathbf{x}}_{t}$ is
\begin{equation}
\underline{\mathbf{P}}_{t|t-1} = \underline{\mathbf{A}}\underline{\mathbf{P}}_{t-1|t-1}\underline{\mathbf{A}}^{H}+
\underline{\mathbf{B}}\underline{\mathbf{Q}}\underline{\mathbf{B}}^{H}=
\begin{bmatrix}\mathbf{P}_{t|t-1} & \widetilde{\mathbf{P}}_{t|t-1} \\
\widetilde{\mathbf{P}}_{t|t-1}^* & \mathbf{P}_{t|t-1}\end{bmatrix},
\end{equation}
where $\mathbf{P}_{t|t-1}$ and $\widetilde{\mathbf{P}}_{t|t-1}$ are the Hermitian and complementary error covariance respectively.
The error covariance matrix $\underline{\mathbf{S}}_{t|t-1}$ for the innovation $\hat{\underline{\mathbf{n}}}_{t|t-1}=\hat{\underline{\mathbf{y}}}_{t|t-1}-\underline{\mathbf{y}}_{t}$ is
\begin{equation}
\underline{\mathbf{S}}_{t|t-1} = \underline{\mathbf{C}}\underline{\mathbf{P}}_{t|t-1}\underline{\mathbf{C}}^{H}+
\underline{\mathbf{R}}=
 \begin{bmatrix} \mathbf{S}_{t|t-1} & \widetilde{\mathbf{S}}_{t|t-1} \\
   \widetilde{\mathbf{S}}_{t|t-1}^* & \mathbf{S}_{t|t-1}^*\end{bmatrix},
   \end{equation}
where $\mathbf{S}_{t|t-1}$ and $\widetilde{\mathbf{S}}_{t|t-1}$ are the Hermitian and complementary innovation covariance respectively.
The normal equation for the Kalman gain is
\begin{equation}
\underline{\mathbf{K}}_{t}\underline{\mathbf{S}}_{t|t-1} = \underline{\mathbf{P}}_{t|t-1}\underline{\mathbf{C}}^{H}.
\end{equation}
Thus the augmented Kalman gain may be written as
\begin{equation}
\underline{\mathbf{K}}_{t} = \underline{\mathbf{P}}_{t|t-1}\underline{\mathbf{C}}^{H}\underline{\mathbf{S}}_{t|t-1}^{-1}
=\begin{bmatrix}\mathbf{K}_{t} & \widetilde{\mathbf{K}}_{t}\\ \widetilde{\mathbf{K}}_{t}^{*} & \mathbf{K}_t^{*}\end{bmatrix}.
\end{equation}
When complementary covariances $\widetilde{\mathbf{P}}_{t|t-1}$ and $\widetilde{\mathbf{S}}_{t|t-1}$ vanish, and when $\mathbf{C}_2$ is zero, we have $\underline{\mathbf{K}}_{t}=diag(\mathbf{K}_t,\mathbf{K}_t^*)$, where $\mathbf{K}_t=\mathbf{P}_{t|t-1}\mathbf{C}_1^H\mathbf{S}_{t|t-1}^{-1}$ is the usual KF. Finally, the WLCKF is
\begin{equation}
\underline{\hat{\mathbf{x}}}_{t|t}=\underline{\hat{\mathbf{x}}}_{t|t-1}+\underline{\mathbf{K}}_t\underline{\hat{\mathbf{n}}}_{t|t-1},
\end{equation}
and the error covariance matrix for $\hat{\underline{\mathbf{e}}}_{t|t}=\hat{\underline{\mathbf{x}}}_{t|t}-\hat{\underline{\mathbf{x}}}_{t}$ is
\begin{eqnarray}\label{wlkf_end}
&&\underline{\mathbf{P}}_{t|t} = \left(\underline{\mathbf{I}}-\underline{\mathbf{K}}_{t}\underline{\mathbf{C}}\right)
\underline{\mathbf{P}}_{t|t-1} =\begin{bmatrix}\mathbf{P}_{t|t} & \widetilde{\mathbf{P}}_{t|t} \\ \widetilde{\mathbf{P}}^{*}_{t|t} & \mathbf{P}^{*}_{t|t}\end{bmatrix}.
\end{eqnarray}
Finally, the WLCKF is implemented by initializing $\hat{\underline{\mathbf{x}}}_{0|0}=\underline{\mathbf{0}}$ and $\underline{\mathbf{P}}_{0|0}=\underline{\mathbf{\Pi}}_{0}$,
and recursively running the procedure (\ref{wlkf_bg})-(\ref{wlkf_end}). This WLCKF can be implemented in complex arithmetic, or it can be inverted for the real KF of the dual channel real model (\ref{ct_kf1})-(\ref{ct_kf2}) by using real to complex connections (\ref{tranf}) and (\ref{agcov}).

\textit{Remark 1}:
In the state and measurement equations (\ref{aug_kf1})-(\ref{matrx}), the new state $\mathbf{x}_t$ depends on $\mathbf{x}_{t-1}$, $\mathbf{x}^*_{t-1}$, $\mathbf{w}_{t-1}$, and $\mathbf{w}^*_{t-1}$. And measurement $\mathbf{y}_t$ depends on $\mathbf{x}_t$, $\mathbf{x}^*_t$, $\mathbf{n}_t$, and $\mathbf{n}^*_t$. For the state and measurement equations of the WLCKF proposed in \cite{Mandic_book}, the new state $\mathbf{x}_t$ depends only on $\mathbf{x}_{t-1}$ and $\mathbf{w}_{t-1}$, and measurement $\mathbf{y}_t$ depends only on $\mathbf{x}_t$ and $\mathbf{n}_t$. Thus the WLCKF in \cite{Mandic_book} can be obtained as a special case of the WLCKF considered here by letting matrices $\mathbf{A}_2$, $\mathbf{B}_2$, and $\mathbf{C}_2$ in (\ref{aug_kf1})-(\ref{matrx}) be zero, or equivalently assuming $\mathbf{E}_{11}=\mathbf{E}_{22}$, $\mathbf{E}_{12}=-\mathbf{E}_{21}$, $\mathbf{F}_{11}=\mathbf{F}_{22}$, $\mathbf{F}_{12}=-\mathbf{F}_{21}$, $\mathbf{G}_{11}=\mathbf{G}_{22}$, and $\mathbf{G}_{12}=-\mathbf{G}_{21}$ in the real channel equations (\ref{ct_kf1})-(\ref{ct_kf2}).

\textit{Remark 2}:
An insightful interpretation of the widely linear KF is that the augmented Kalman gain determines the WLMMSE estimator of the prediction error $\hat{\mathbf{e}}_{t|t-1}=\mathbf{x}_{t} - \hat{\mathbf{x}}_{t|t-1}$ from the innovation $\hat{\mathbf{n}}_{t|t-1}=\mathbf{y}_{t}-\hat{\mathbf{y}}_{t|t-1}$. Thus the widely linear complex KF reduces to the linear complex KF if and only if $\hat{\mathbf{e}}_{t|t-1}-\mathbf{K}_t\hat{\mathbf{n}}_{t|t-1}$ is orthogonal to $\hat{\mathbf{n}}_{t|t-1}^*$, where $\mathbf{K}_t=\mathbf{P}_{t-1|t-1}
\mathbf{C}^{H}\mathbf{S}_{t|t-1}^{-1}$. That is,
\begin{equation}\begin{split}
E(\hat{\mathbf{e}}_{t|t-1}-\mathbf{K}_t\hat{\mathbf{n}}_{t|t-1})\hat{\mathbf{n}}_{t|t-1}^T &=E\hat{\mathbf{e}}_{t|t-1}(\mathbf{C}\hat{\mathbf{e}}_{t|t-1}+\mathbf{n}_t)^T
-\mathbf{K}_tE\hat{\mathbf{n}}_{t|t-1}\hat{\mathbf{n}}_{t|t-1}^T \\
&=\widetilde{\mathbf{P}}_{t|t-1}\mathbf{C}^{T}-\mathbf{P}_{t-1|t-1}
\mathbf{C}^{H}\mathbf{S}_{t|t-1}^{-1}\widetilde{\mathbf{S}}_{t|t-1} \\ &= \mathbf{0} . \nonumber\
\end{split}\end{equation}
One special case is that the error covariance of the predictor $\hat{\mathbf{e}}_{t|t-1}$ is proper, $\widetilde{\mathbf{P}}_{t|t-1} = \mathbf{0}$, and the innovation $\hat{\mathbf{n}}_{t|t-1}$ is proper, $\widetilde{\mathbf{S}}_{t|t-1} = \mathbf{0}$. This is true when $\hat{\mathbf{e}}_{t-1|t-1}$, $\mathbf{w}_{t}$, and $\mathbf{n}_{t}$ are all proper. Another special case is that $\hat{\mathbf{n}}_{t|t-1}$ is maximally improper, i.e., $\hat{\mathbf{n}}_{t|t-1}=\alpha\hat{\mathbf{n}}_{t|t-1}^*$ with probability 1 for constant scalar $|\alpha|=1$. This is irrespective of whether $\hat{\mathbf{e}}_{t|t-1}$ is improper. For instance, assume
$\mathbf{y}_t=\textrm{Re}(\mathbf{x}_t)+\mathbf{n}_t$ with $\mathbf{n}_t$ real. We know that $\mathbf{y}_t$ is a noisy widely linear transformation of $\mathbf{x}_t$ and $\hat{\mathbf{y}}_{t|t-1}=\textrm{Re}(\hat{\mathbf{x}}_{t|t-1})$, meaning the innovation $\hat{\mathbf{n}}_{t|t-1}$ is real and hence maximally improper. One can also readily see that no widely linear processing is needed because $\mathbf{y}_t$ is only dependent on
the real part of $\mathbf{x}_t$.

\section{Performance Comparison between WLCKF and CKF}
Let's suppose the state and measurement equations for a complex state $x_t$ are
\begin{equation}\label{ex2_eq1}
x_t = a_{t-1}x_{t-1}+b_{t-1}w_{t-1},\ t=1,2,...,
\end{equation}
\begin{equation}\label{ex2_eq2}
y_t = c_tx_t + n_t,\ t= 0,1,...,
\end{equation}
where $a_t,b_t,c_t\in\mathbb{C}$ for all $t\geq0$. The augmented matrices are $\underline{\mathbf{A}}_t=diag(a_t,a_t^*)$, $\underline{\mathbf{B}}_t=diag(b_t,b_t^*)$, and $\underline{\mathbf{C}}_t=diag(c_t,c_t^*)$. Then the recursion for the $2$ by $2$ augmented covariance matrix $\underline{\mathbf{P}}_{t|t}$ is
\begin{equation}\begin{split}
\underline{\mathbf{P}}_{t|t}
& = (\underline{\mathbf{P}}_{t|t-1}^{-1}+\underline{\mathbf{C}}_{t}^H\underline{\mathbf{R}}^{-1}\underline{\mathbf{C}}_{t})^{-1} \\
&=\left[(|a_t|^2\underline{\mathbf{P}}_{t-1|t-1}+|b_t|^2\underline{\mathbf{Q}})^{-1}
+|c_t|^2\underline{\mathbf{R}}^{-1}\right]^{-1},\ t = 1,2,....
\end{split}\end{equation}
Thus the performance of the WLCKF is determined by the impropriety of the initial state $x_0$ through $\underline{\boldsymbol{\Pi}}_0$, the driving noise $w_t$ through $\underline{\mathbf{Q}}$, and the measurement noise $n_t$ through $\underline{\mathbf{R}}$. In the following we show that for some special distributions of state and noises, the WLCKF produces smaller MSE than the CKF.

\textit{Case 1: }$x_0$ is improper, $w_t$ and $n_t$ are proper.

Suppose $\underline{\mathbf{P}}_{0|0}=\left(\begin{smallmatrix}P_{0|0}&\widetilde{P}_{0|0}\\ \widetilde{P}_{0|0}^*&P_{0|0}\end{smallmatrix} \right)$, $\underline{\mathbf{Q}}=N_1\underline{\mathbf{I}}$, and $\underline{\mathbf{R}}=N_2\underline{\mathbf{I}}$.
Assume $\underline{\mathbf{P}}_{0|0}$ has eigenvalues $\{\lambda_{1}^{0},\lambda_{2}^{0}\}$. Given the eigenvalues $\{\lambda_{1}^{t-1},\lambda_{2}^{t-1}\}$ of matrix $\underline{\mathbf{P}}_{t-1|t-1}$, the eigenvalues of $\underline{\mathbf{P}}_{t|t}$ are
\begin{equation}
\lambda_i^t=g_t(\lambda_{i}^{t-1}),\ \ i=1,2, \nonumber\
\end{equation}
where the function $g_t$ is given by
\begin{equation}
g_t(\lambda) = \frac{N_2(|a_t|^2\lambda+|b_t|^2N_1)}{|c_t|^2(|a_t|^2\lambda+|b_t|^2N_1)+N_2},\ \ t = 1,2,... \nonumber\
\end{equation}
Thus the eigenvalues $\{\lambda_{1}^t,\lambda_{2}^t\}$ may be conveniently expressed as the function recursion
\begin{equation}
\begin{split}
\lambda_i^t = g_t\circ g_{t-1}\circ\cdot\cdot\cdot\circ g_{1}(\lambda_{i}^0) \triangleq q_t(\lambda_{i}^0) \nonumber\
\end{split}
\end{equation}
Observe that $g_t$ is an increasing concave function w.r.t $\lambda$ for each $t$. Thus we conclude that $q_t$ is concave for each $t$. Next we want to compute the widely linear minimum mean squared error (WLMMSE) at the $t$-th iteration for the WLCKF. This may be written
\begin{equation}\begin{split}
\xi^\textrm{WL}_t = E\parallel \hat{e}_{t|t}\parallel^2 = \frac{1}{2}\textrm{tr}(\underline{\mathbf{P}}_{t|t}) =\frac{1}{2}(q_t(\lambda_{1}^0)+q_t(\lambda_{2}^0)).
\end{split}\label{WLMMSE}\end{equation}
Note the initial \emph{scalar} Hermitian covariance is $P_{0|0}$. Then the $t$-th LMMSE for the CKF is
\begin{equation}
\xi^\textrm{L}_t = q_t(P_{0|0}). \nonumber\
\end{equation}
To achieve the maximum performance improvement of the WLCKF over the CKF for the special case discussed here, we shall minimize $\xi^\textrm{WL}_t$ with fixed $P_{0|0}$ and variable $\widetilde{P}_{0|0}$. It can be seen that at each $t$, $\xi^\textrm{WL}_t$ is a Schur-concave function w.r.t all $\lambda_{i}^0$. Since $\lambda_1^0+\lambda_2^0\leq2P_{0|0}$ \cite{book_PJ&LL}, the minimum is achieved when
\begin{equation}
[\lambda_{1}^0 \ \lambda_{2}^0]=[2P_{0|0} \ 0].\label{lmd_opt}
\end{equation}
Substituting (\ref{lmd_opt}) into (\ref{WLMMSE}), we have the minimum $\xi^\textrm{WL}_t$:
\begin{equation}
\min{\xi^\textrm{WL}_t} = \frac{1}{2}\left(q_t(2P_{0|0})+q_t(0)\right) \nonumber\
\end{equation}
The ratio of $\min{\xi^\textrm{WL}_t}$ to $\xi^\textrm{L}_t$ is:
\begin{equation}\begin{split}\theta_t=\frac{\min{\xi^\textrm{WL}_t}}{\xi^\textrm{SL}_t}\end{split}=
\frac{q_t(2P_{0|0})+q_t(0)}{2q_t(P_{0|0})}\nonumber\
\end{equation}
It's obvious that $\frac{1}{2}\leq\theta_t\leq1$. This is because $q_t$ is concave, and $q_t(2P_{0|0})+q_t(0)\leq2q_t(P_{0|0})$ for any $P_{0|0}$. Also $q_t(2P_{0|0})\geq q_t(P_{0|0})$ for any $P_{0|0}$ and $q_t(0)\geq 0$. Actually the condition for achieving the lower bound is $N_1\ll N_2\ll 1$. This coincides with the MMSE analysis in \cite{book_PJ&LL}.
\begin{figure}
\subfigure[]{
\begin{minipage}[b]{0.3\textwidth}
\centering
\includegraphics[width=2.1in]{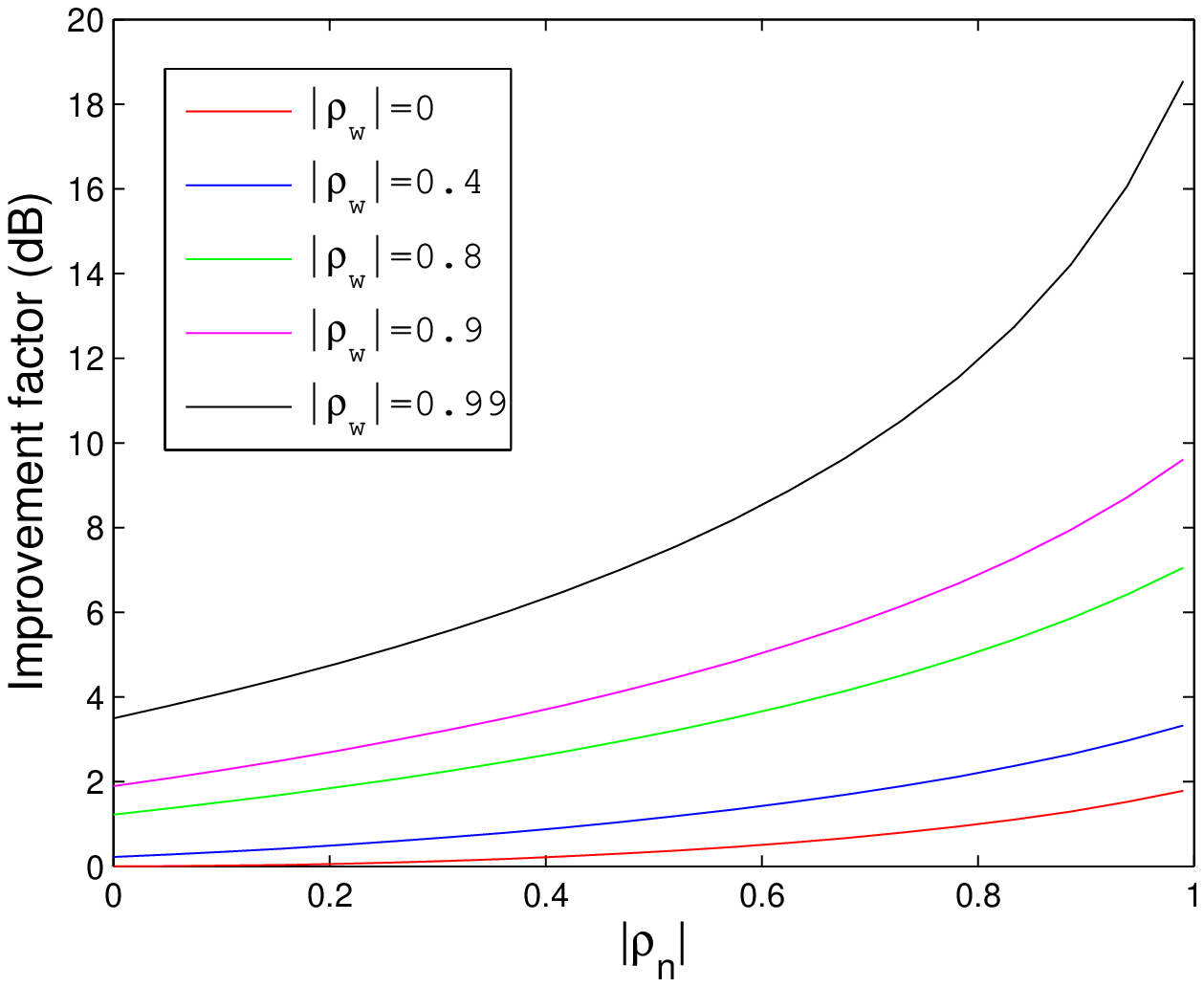}
\end{minipage}}
\subfigure[]{
\begin{minipage}[b]{0.3\textwidth}
\centering
\includegraphics[width=2.1in]{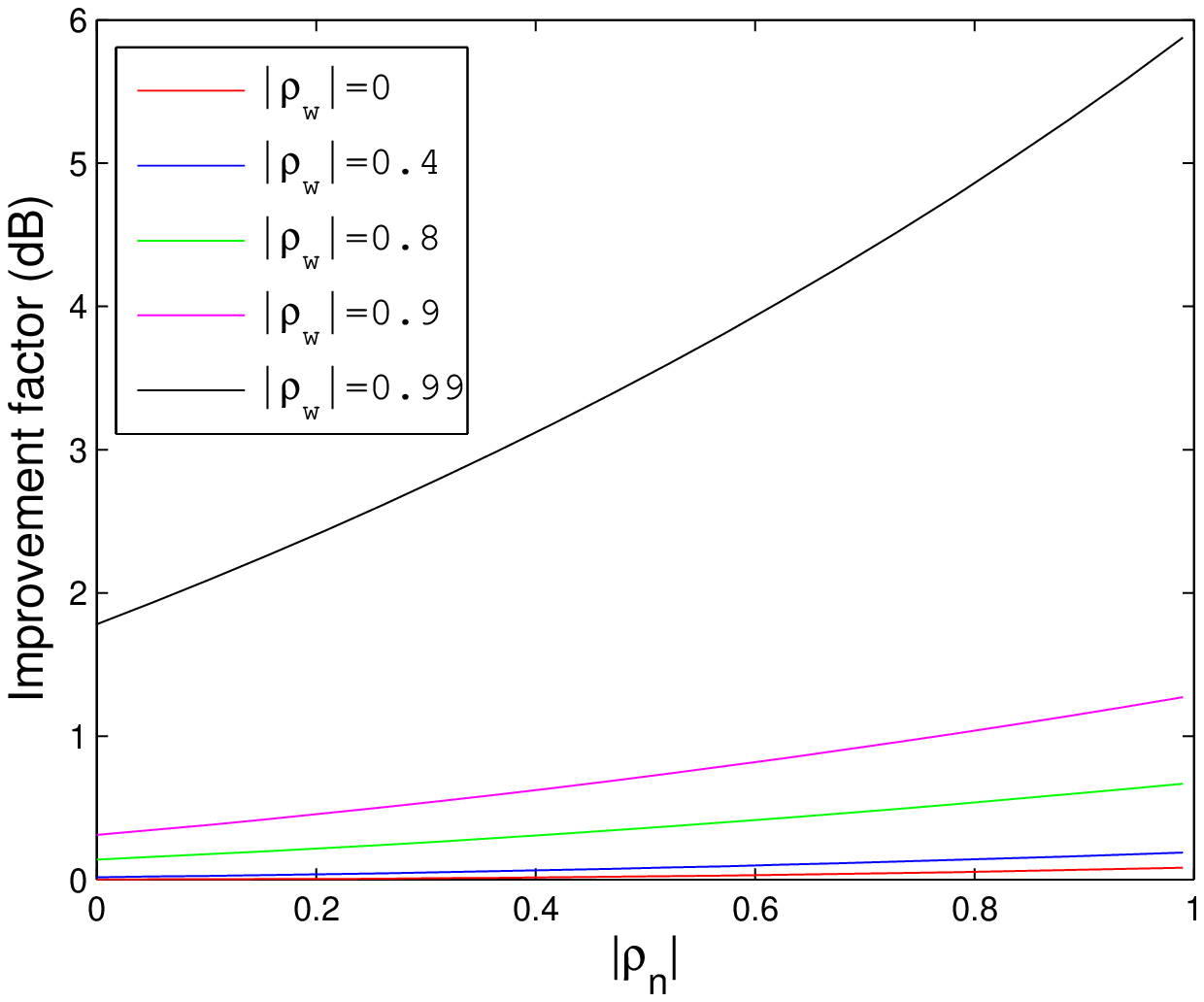}
\end{minipage}}
\subfigure[]{
\begin{minipage}[b]{0.30\textwidth}
\centering
\includegraphics[width=2.1in]{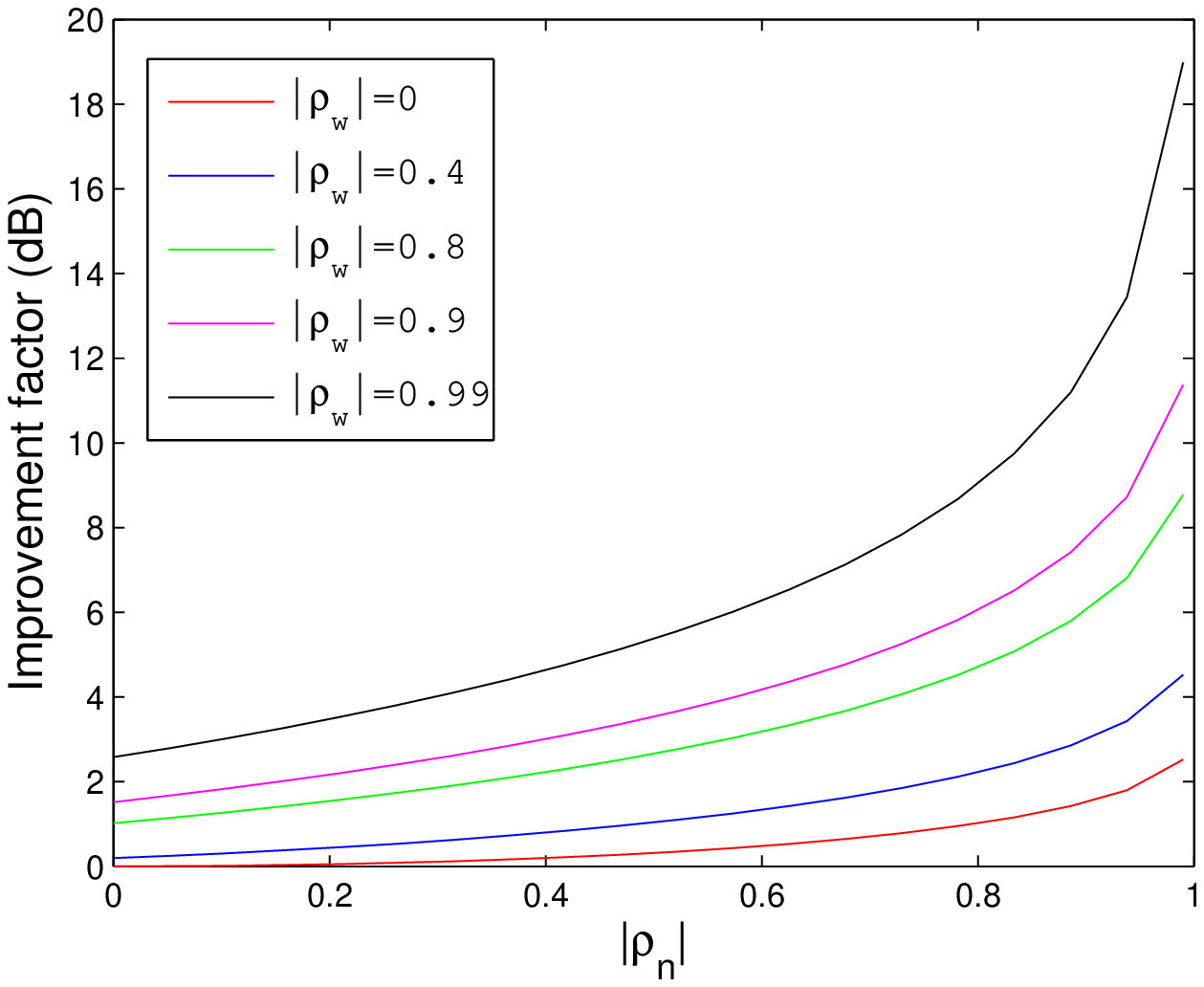}
\end{minipage}}
\caption{MSE performance improvement of the WLCKF over the CKF. (a) $N_1=-20\textrm{dB}$, $N_2 = -20\textrm{dB}$. (b) $N_1 = -20\textrm{dB}$, $N_2=-40\textrm{dB}$. (c) $N_1 = -40\textrm{dB}$, $N_2=-20\textrm{dB}$.}
\end{figure}

\textit{Case 2: }$x_0$ is proper, $w_t$ and $n_t$ are improper.

In this case we have $\underline{\mathbf{P}}_{0|0}=P_{0|0}\underline{\mathbf{I}}$, $\underline{\mathbf{Q}}=N_1\left(\begin{smallmatrix}1&\rho_{w}\\ \rho_{w}^*&1\end{smallmatrix} \right)$, and $ \underline{\mathbf{R}}=N_2\left(\begin{smallmatrix}1&\rho_{n}\\ \rho_{n}^*&1\end{smallmatrix} \right)$,
where $\rho_{w}$ is the \textit{complex correlation coefficient} between $w_t$ and $w_t^*$, and $\rho_{n}$ is the complex correlation coefficient between $n_t$ and $n_t^*$. These determine the level of impropriety. We can show that $0\leq |\rho_{w}|,|\rho_{n}| \leq 1$. Fig 1 plots the performance improvement of the WLCKF over the CKF at different level of impropriety of $w_t$ and $n_t$. We choose $P_{0|0}=1$, $a = b = c = 1$. The performance improvement is defined by ratio between the convergent MSE of the CKF over that of the WLCKF. As Fig. 1 illustrates, the performance improvement is monotone in $|\rho_w|$ for fixed $|\rho_n|$, and monotone in $|\rho_n|$ for fixed $|\rho_w|$.

\section{Unscented Widely Linear Kalman Filter}
In this section we consider the following nonlinear model for dual real channel state and measurement evolution:
\begin{equation}
\mathbf{z}_t = [\mathbf{u}_t \ \mathbf{v}_t]^T = \mathbf{f}_{t-1}(\mathbf{z}_{t-1},\boldsymbol{\omega}_{t-1}) = \mathbf{f}_{t-1}([\mathbf{u}_{t-1} \ \mathbf{v}_{t-1}]^T,[\boldsymbol{\mu}_t \ \boldsymbol{\sigma}_t]^T), t = 1, 2, ...,
\end{equation}
\begin{equation}
\boldsymbol{\psi}_t = [\boldsymbol{\xi}_t\ \boldsymbol{\kappa}_t]^T = \mathbf{h}_{t}(\mathbf{z}_t,\boldsymbol{\eta}_t) = \mathbf{h}_{t}([\mathbf{u}_t \ \mathbf{v}_t]^T,[\boldsymbol{\rho}_t \ \boldsymbol{\phi}_t]^T), t = 0,1,...,
\end{equation}
where $\mathbf{f}_{t-1}$ and $\mathbf{h}_{t}$ are time varying nonlinear transformations, and the notation for states and noises is identical with the model equations (\ref{ct_kf1})-(\ref{ct_kf2}). Then the induced complex model equations are
\begin{equation}
\mathbf{x}_t = \widetilde{\mathbf{f}}_{t-1}(\mathbf{x}_{t-1},\mathbf{w}_{t-1}),
\end{equation}
\begin{equation}
\mathbf{y}_t = \widetilde{\mathbf{h}}_{t}(\mathbf{x}_t,\mathbf{n}_t).
\end{equation}
It can be seen that for all $t$, $\widetilde{\mathbf{f}}_{t-1}$ and $\widetilde{\mathbf{h}}_{t}$ are not widely linear transformations. Thus the WLCKF developed in section III cannot be directly utilized. For such a model, the extended WLCKF is proposed in \cite{Mandic_book, Mandic_conf} to exploit the impropriety of complex states and noises. However, the major defect of the EWLCKF is that the posterior means and covariances are accurate only to the first order in a Taylor expansion. A conventional Unscented KF uses the unscented transformation (UT) to generate a fixed set of sigma points to represent the distribution of a random variable \cite{UKF_Julier1}. After propagating sigma points through nonlinearities, the estimated posterior mean and covariance are precise at least to second order in a Taylor expansion. Motivated by the power of UKF, in this section we present a novel paradigm for constructing UWLCKFs. Our UWLCKFs use \textit{modified} sigma points which preserve the Hermitian \textit{and} complementary covariances of states and noises, while the UWLCKFs proposed in \cite{Mandic_book} use sigma points which only preserve the Hermitian covariances of states and noises.

Compose complex random states and noises into a complex vector $\mathbf{s}^T=[\mathbf{x}^T\  \mathbf{w}^T \ \mathbf{n}^T]$. Suppose the augmented mean and covariance of $\mathbf{s}$ are
\begin{equation}\label{mean&cov}
\underline{\boldsymbol{\mu}}_{s}^T = [\boldsymbol{\mu}_s^T\ \boldsymbol{\mu}_s^H],\
\underline{\mathbf{R}}_{ss} =\begin{bmatrix}\mathbf{R}_{ss} & \widetilde{\mathbf{R}}_{ss} \\ \widetilde{\mathbf{R}}_{ss}^* & \mathbf{R}_{ss}^*\end{bmatrix}.
\end{equation}
In \cite{Mandic_book} the authors proposed complex sigma points of $\mathbf{s}$ which are constructed from moments $\boldsymbol{\mu}_s$ and $\mathbf{R}_{ss}$. Thus these sigma points only carry $\boldsymbol{\mu}_s$ and $\mathbf{R}_{ss}$, but not $\widetilde{\mathbf{R}}_{ss}$. In fact, there may be multiple ways to generate sigma points for the augmented random vector $\underline{\mathbf{s}}$ which carry both $\underline{\boldsymbol{\mu}}_{s}$ and $\underline{\mathbf{R}}_{ss}$. But a hidden restriction
imposed here is that these sigma points should be augmented vectors. Otherwise they cannot be propagated
through the UWLCKF. One approach is to start with sigma points of the corresponding composite real random vector $\boldsymbol{\zeta}^T = [\mathbf{u}^T\ \boldsymbol{\mu}^T\ \boldsymbol{\rho}^T\ \mathbf{v}^T\ \boldsymbol{\sigma}^T\ \boldsymbol{\phi}^T]$. The first and second moments of $\boldsymbol{\zeta}$ are
\begin{equation}
\boldsymbol{\mu}_{\zeta} = \frac{1}{2}\mathbf{T}^{-1}\underline{\boldsymbol{\mu}}_{s},\ \mathbf{R}_{\zeta\zeta} = \frac{1}{4}\mathbf{T}^H\underline{\mathbf{R}}_{ss}\mathbf{T} .
\end{equation}
Using a Cholesky decomposition the composite covariance matrix $\mathbf{R}_{\zeta\zeta}$ may be factored as
\begin{equation}\label{chol}
\mathbf{R}_{\zeta\zeta} = \mathbf{BB}^T \nonumber\
\end{equation}
Denote the vector $\mathbf{b}_k$ as the $k$-th column of matrix $\mathbf{B}$ for $k=1,2,...,2N$. Then the sigma points $\{\boldsymbol{\mathcal{Z}}_k\}$ of $\boldsymbol{\zeta}$ are \cite{UKF_Julier1}
\begin{eqnarray}
&&\boldsymbol{\mathcal{Z}}_0 = \boldsymbol{\mu}_\zeta,\ k=0,\\ \nonumber\
&&\boldsymbol{\mathcal{Z}}_k = \boldsymbol{\mu}_\zeta + \sqrt{2N+\lambda}\mathbf{b}_k,\ k=1,...,2N, \\ \nonumber\
&&\boldsymbol{\mathcal{Z}}_k = \boldsymbol{\mu}_\zeta - \sqrt{2N+\lambda}\mathbf{b}_{k-2N},\ k=2N+1,...,4N,
\end{eqnarray}
corresponding to the mean weights $\{W_m(k)\}_{k=0}^{4N}$ and covariances weights $\{W_c(k)\}_{k=0}^{4N}$ defined in \cite{UKF_Julier1}.
Define a set of augmented vectors $\{\underline{\boldsymbol{\mathcal{X}}}_k\}$ as
\begin{eqnarray}\label{sigma_pt}
&&\underline{\boldsymbol{\mathcal{X}}}_k =\begin{bmatrix}\boldsymbol{\mathcal{X}}_k \\ \boldsymbol{\mathcal{X}}_k^*\end{bmatrix}= \mathbf{T}\boldsymbol{\mathcal{Z}}_k=\begin{cases}\underline{\boldsymbol{\mu}}_s, & k=0,
\\ \underline{\boldsymbol{\mu}}_s + \sqrt{2N+\lambda}\mathbf{T}\mathbf{b}_k,& k=1,...,2N,
\\ \underline{\boldsymbol{\mu}}_s - \sqrt{2N+\lambda}\mathbf{T}\mathbf{b}_{k-2N},& k=2N+1,...,4N.
\end{cases}
\end{eqnarray}
We can show that all the $\underline{\boldsymbol{\mathcal{X}}}_k$ compose the sigma points of the augmented vector $\underline{\mathbf{s}}$, since $\underline{\boldsymbol{\mathcal{X}}}_0 = \underline{\boldsymbol{\mu}}_s$ and
\begin{equation}\begin{split}\label{sigma_cov}
\underline{\mathbf{R}}_{ss}&=\mathbf{T}\mathbf{R}_{\zeta\zeta}\mathbf{T}^H \\ \nonumber\
&=\begin{bmatrix}\mathbf{T}\mathbf{b}_1& \mathbf{T}\mathbf{b}_2 & \cdots &\mathbf{T}\mathbf{b}_{2N}\end{bmatrix} \begin{bmatrix}\mathbf{T}\mathbf{b}_1& \mathbf{T}\mathbf{b}_2 & \cdots &\mathbf{T}\mathbf{b}_{2N}\end{bmatrix}^H. \nonumber\
\end{split}\end{equation}
Therefore we have obtained the sigma points $\{\underline{\boldsymbol{\mathcal{X}}}_k\}$ of $\underline{\mathbf{s}}$ w.r.t weights $\{W_m(k),W_c(k)\}$ from widely linear transformation of the real composite sigma points $\{\boldsymbol{\mathcal{Z}}_k\}$ of $\boldsymbol{\zeta}$ w.r.t weights $\{W_m(k),W_c(k)\}$. Note that each sigma point $\underline{\boldsymbol{\mathcal{X}}}_k$ is an augmented vector. Thus it follows that the complex set $\{\boldsymbol{\mathcal{X}}_k\}$, generated by extracting the top halves of $\{\underline{\boldsymbol{\mathcal{X}}}_k\}$, is sufficient to capture both first and second order statistical information of the augmented random vector $\underline{\mathbf{s}}$. We call $\{\boldsymbol{\mathcal{X}}_k\}$ the \emph{modified} sigma points of $\mathbf{s}$. The impact of these modified sigma points is that $\{\boldsymbol{\mathcal{X}}_k\}$ preserves not only mean $\boldsymbol{\mu}_s$ and Hermitian covariance $\mathbf{R}_{ss}$, but also complementary covariance $\widetilde{\mathbf{R}}_{ss}$.


\emph{Example 1 (Phase Demodulation Problem):}$\ $ Consider a scalar real random phase $\theta_t$ that is updated as
\begin{equation}
\theta_t = a\theta_{t-1}+bw_{t-1},\ t=1,2,...,
\end{equation}
where $w_t$ is a real driving noise. So, the phase is real, and it evolves or jitters according to a first-order Markov sequence. The measurement in a quadrature demodulator is a noisy complex signal modulated by $\theta_t$:
\begin{equation}
y_t = e^{i\theta_t}+n_t,\ t=0,1,... \end{equation}
where each $n_t$ is assumed to be a zero mean, scalar \emph{complex} Gaussian random variable \cite{book_PJ&LL} with Hermitian variance $R$ and complementary variance $\widetilde{R}$.
The complex correlation coefficient between $n_t$ and $n_t^*$ is
$\rho = \frac{\widetilde{R}}{R}$
which describes the impropriety of $n_t$. 
The signal-to-noise ratio at the receiver is SNR$\ = R^{-1}$. In simulation we set $a=0.98$, $b=0.05$. Each $w_t$ is a standard mean zero and variance one Gaussian real random variable, independent of all others.

Fig. 2(a) draws the outputs of the UWLCKF over time at SNR$\ =30$dB and $|\rho|=0.5$. The UWLCKF is constructed according to Algorithm 3. The widely linear Kalman gain for the UWLCKF is a $2$ by $2$ matrix and the estimate $\hat{\theta}_{t|t}$ is always real. It can be observed that for most iterations, the estimate $\hat{\theta}_{t|t}$ is close to the phase $\theta_t$. Also the true $\theta_t$ is almost confined by the envelope $\hat{\theta}_{t|t}\pm \sqrt{P_{t|t}}$.

Fig. 2(b) compares the performances of the UWLCKF that accounts for the impropriety of the noise, and the UKF that assumes the noise to be proper. Unlike the UWLCKF above, the UKF estimates $\theta_t$ from a real $2$ by $1$ measurement vector consisting of the real and imaginary part of $y_t$ collected from dual channels. At each iteration the UKF produces sigma points from the real mean vector $[\hat{\theta}_{t|t}\  \mu_w\ \mu_u \ \mu_v]^T=[\hat{\theta}_{t|t}\ 0\ 0 \ 0]^T$ and covariance matrix $\mathbf{M}=diag(P_{t|t}, 1, R_u, R_v)$, and it has a $1$ by $2$ Kalman gain vector. The complex correlation coefficient is
$|\rho|=0.7$. Define the normalized squared error as $\xi=||\mathbf{e}||^2_2/||\boldsymbol{\theta}||^2_2$, where $\boldsymbol{\theta}$ and $\mathbf{e}$ are vectors consisting of phases and estimation errors in 500 iterations respectively. In the plot each $\xi$ is computed by averaging 1000 Monte-Carlo simulations. It can be seen that in the low-medium SNR regime, UWLCKF requires about 2dB less SNR than the UKF.

Fig. 2(c) shows the performance improvement of the UWLCKF over the UKF vs the noise impropriety $|\rho|$ at different SNRs. We use the factor $r=\xi_{\textrm{UKF}}/\xi_{\textrm{UWLCKF}}$ to evaluate the advantage of UWLCKF. The normalized squared error $\xi_{\textrm{UKF}}$ and $\xi_{\textrm{UWLCKF}}$ are defined as above. Each $r$ is computed by averaging 1000 Monte-Carlo simulations. For $|\rho|\geq 0.8$, the gain $r\geq 2$.
%
%
\begin{figure}
\subfigure[]{
\begin{minipage}[b]{0.3\textwidth}
\centering
\includegraphics[width=2.1in]{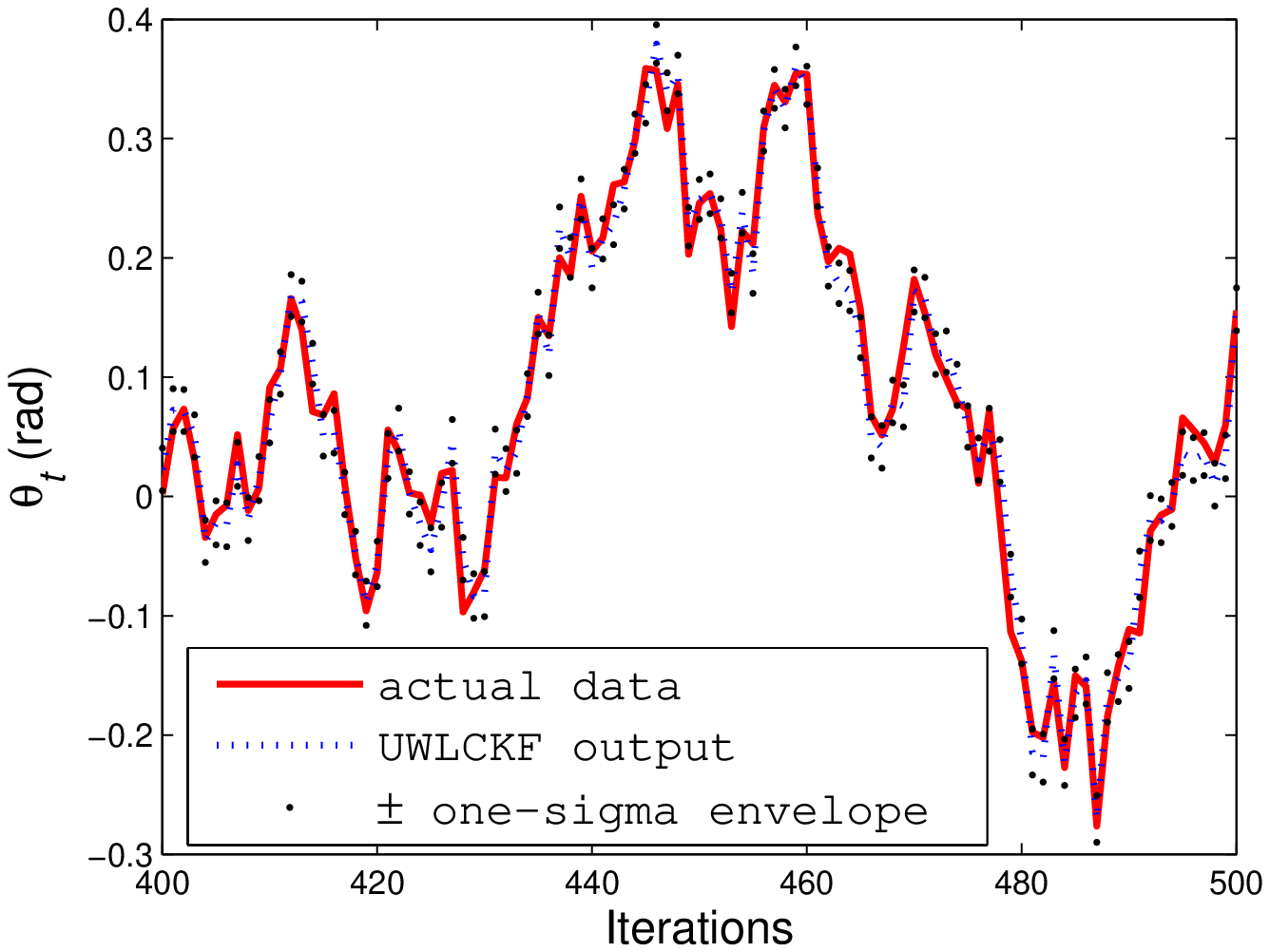}
\end{minipage}}
\subfigure[]{
\begin{minipage}[b]{0.3\textwidth}
\centering
\includegraphics[width=2.1in]{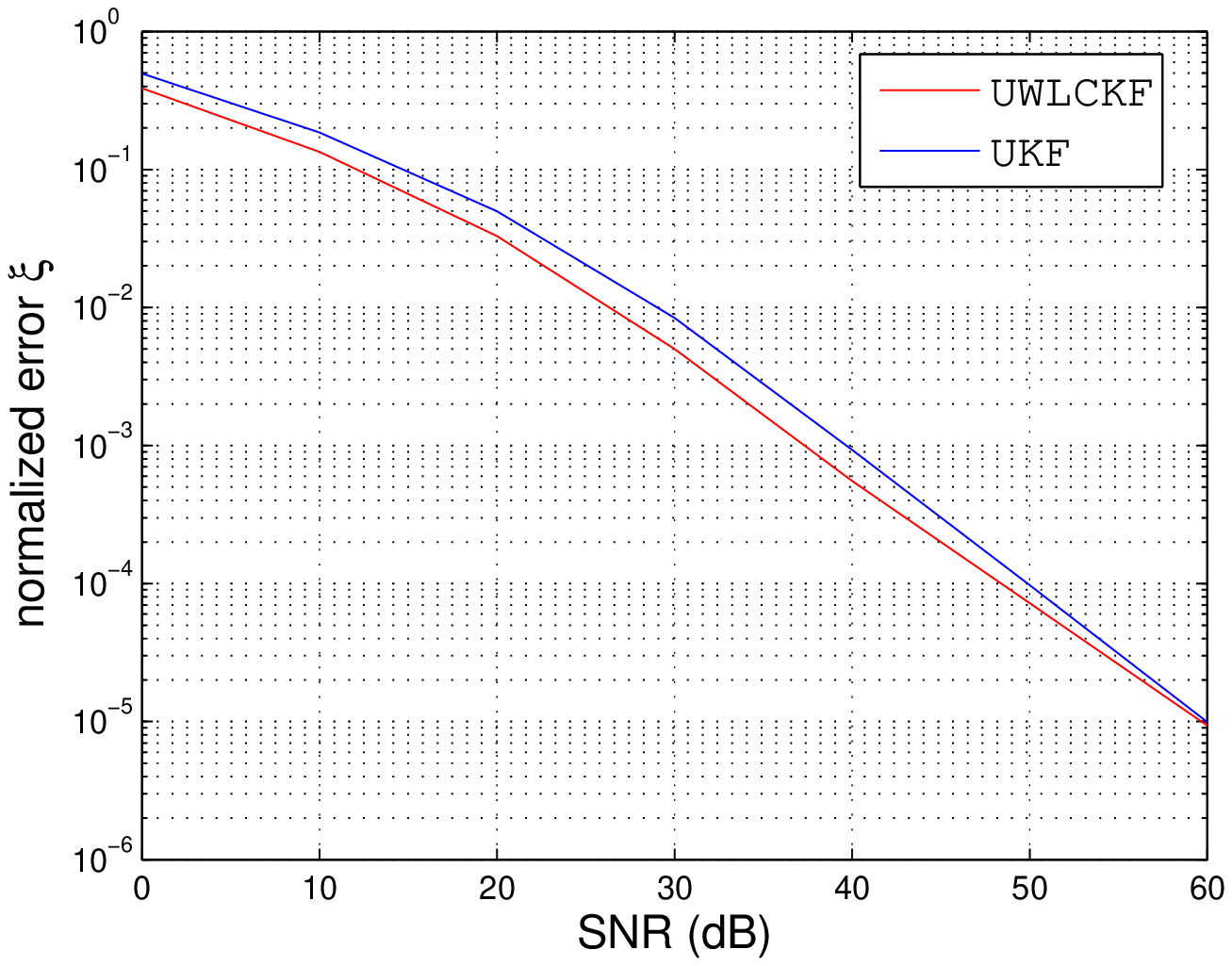}
\end{minipage}}
\subfigure[]{
\begin{minipage}[b]{0.30\textwidth}
\centering
\includegraphics[width=2.1in]{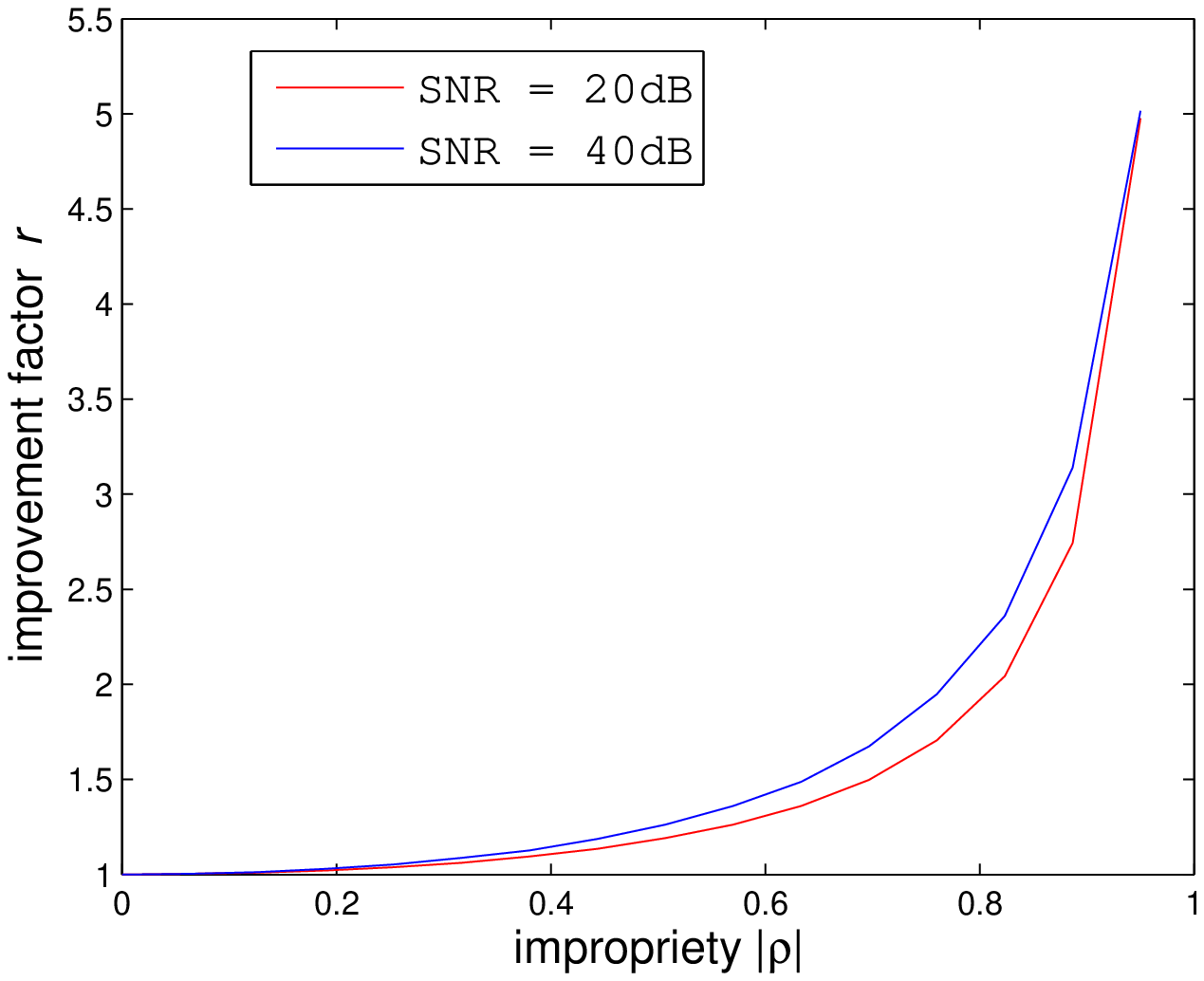}
\end{minipage}}
\caption{Comparison between UWLCKF and UKF. (a) Phase estimated by UWLCKF at each iteration, SNR = 30dB, $|\rho|=0.5$. (b) Normalized estimation error $\xi$ of UWLCKF and UKF vs SNRs, $|\rho|=0.7$. (c) Performance improvement $r$ of UWLCKF over UKF vs impropriety of $n_t$.}
\end{figure}

\section{Conclusion}
In this paper we have designed widely linear and unscented WL complex Kalman filters for complex noisy dynamical systems with improper states and noises. We show that WLCKFs may significantly improve on the performance of a CKF that ignores corresponding covariance. A simulation for real phase demodulation shows how an UWLCKF produces real estimates from complex baseband measurements and shows the improvement of its performance over an unscented complex KF that assumes proper states and noises.
\bibliographystyle{IEEEtran}
\bibliography{reference}
\end{document}